\documentclass[12pt, oneside]{amsart}
\pdfoutput=1
\usepackage{geometry}
\usepackage{lmodern}
\allowdisplaybreaks
\usepackage{amsmath}
\usepackage{amssymb}
\usepackage{amsthm}
\usepackage{color}
\usepackage[usenames,dvipsnames]{xcolor}
\usepackage{graphicx}
\usepackage{threeparttable}
\usepackage{subfig}
\usepackage{soul}
\usepackage[countmax]{subfloat}
\usepackage{float}
\usepackage{rotfloat}
\usepackage{comment}
\usepackage{setspace}
\usepackage{esint}
\usepackage{tikz}
\usepackage{multicol}
\usepackage{lscape}
\usepackage{mathtools}
\usepackage[authoryear]{natbib}
\definecolor{MyDarkBlue}{rgb}{0,0.08,0.45}
\usepackage[unicode=true,pdfusetitle, bookmarks=false,bookmarksnumbered=false,bookmarksopen=false, breaklinks=false,pdfborder={0 0 1},colorlinks=true, linkcolor = MyDarkBlue, citecolor = MyDarkBlue]{hyperref}
\usepackage{breakurl}

\PassOptionsToPackage{normalem}{ulem}
\usepackage{ulem}

\makeatletter

\numberwithin{equation}{section}

\providecommand{\E}{\mathrm{E}}

\makeatother
\begin{document}

\title[]{Experimenting with Networks}

\author{Arun G. Chandrasekhar$^{\ddagger,\S}$}
\author{Matthew O. Jackson$^{\ddagger, \star}$ }
\date{June 2025}

\thanks{$^{\ddagger}$Stanford, Department of Economics}
\thanks{$^{\S}$J-PAL}
\thanks{$^{\star}$Santa Fe Institute}

\begin{abstract}

We provide an overview of methods for designing and implementing experiments (field, lab, hybrid, and natural) when there are networks of interactions between subjects.

Keywords:  experiments, networks, social networks, field experiments, lab experiments, natural experiments

{\sl JEL codes}: D85, O1, C90, C91, C92, C93
\end{abstract}

\thispagestyle{empty}

\maketitle

Understanding how an economic or social environment affects behavior---and how that depends on network interaction patterns among agents themselves---is vital to both understanding human behavior and designing policies to improve welfare when there are inefficiencies. 
Experimentation in its various forms is the main tool that allows for causal inference (e.g., \cite{duflo2007using,falk2009lab,croson2010science,montgomery2017design,baldassarri2017field,list2025}).

Experimentation is challenging, as a researcher often has only partial control over all variables that influence outcomes.
This becomes even more challenging when trying to causally understand the determinants of behaviors on networks where there are externalities and peer interactions both in treatment selection and behaviors.
It is important to emphasize that much of the research into individual behavior has been in settings in which networked interactions matter, but are ignored.  It is actually hard to find examples of choices (including participation in a treatment) that are not peer-influenced.   Ignoring the network aspect can lead to substantially biased conclusions.

In this chapter we examine designing, conducting, and analyzing experiments in settings in which networked interactions play a role in determining people's behaviors and welfare.  This includes both how networks influence behaviors and, conversely, how settings and behaviors influence network formation.

Let us begin by mentioning some key challenges that are faced in doing research in the presence of network effects.

First, there are multiple forms of endogeneity. By definition, an individual’s outcomes are co-determined with others’ and therefore directly or indirectly depend on others’ decisions. Given that behaviors are mutually determined, they are statistically dependent.
Moreover, these decisions can be jointly sensitive to contextual features. For example, endogenous things like culture, norms, and beliefs can change the nature of strategic interactions.
In addition, the actual patterns of interaction can be endogenously determined by preferences and/or behaviors.  People often prefer to interact with others who have similar preferences and behaviors, which endogenizes the network and feedback effects.

Second, there are data challenges.   Adequate data on the network patterns of interactions can be difficult to gather.  Moreover, a priori, it is can be unclear what the appropriate relationships are that might influence behaviors  \citep{manski1993,chandrasekhar2015econometrics}.  The network tracks the potential interference/influence patterns and without good measurement of it, one may be very limited in what conclusions can be drawn.  The bottom line is that in order to do causal analysis, one often needs to have an accurate picture of the network governing interactions, but this may be difficult to ascertain both because it is hard to know all of the possible interaction channels, and because it can be hard to measure the network even if one knows what one wants to measure.

Against this backdrop, and also {\sl because} of these challenges,  experiments are very valuable tools in the study of social and economic networks:
\begin{itemize}
\item First, experiments allow one to at least partially control or generate variation in interaction patterns or the things that drive those patterns. Given the interdependence in behaviors, without some form of controlled or random assignment it can be hopeless to make credible statements.
\item Second, even if one cannot control the network itself, generating clean variation in treatment at least allows networks to be studied through the language of heterogeneous effects if one has a good picture of the relevant network.
\item Third, if one can control various features of the environment (e.g., who is exposed to whom), then one can affect and control network formation, which is often an object of interest itself.
\end{itemize}

There are three classes of experiments at our disposal: lab, field, and natural.
Lab experiments allow for great control over interaction patterns, structure of information sets, as well as the generation of panel data holding the network fixed  over time (e.g., see chapters 1, 6, and 11). However, by definition, estimates generated from lab experiments come from an isolated setting. They can be powerful devices for understanding human behavior and social interactions, but may rely on limited settings both in scale and scope.  For instance, sometimes, though not always, labs use lower stakes and artificial and small networks.
Field experiments allow us to dive into more complex, ``real world'' settings, but limit the ability not only to control but even to know the shape of interaction and social influence. This could both be because the definition of a link is less clear and the scope for spillovers less controlled but also because these typically must be done in the cross-section; and can require greater assumptions on the nature of spillovers.
In principle, when causality is credible, natural experiments offer the most direct observation of behavior in the wild, while providing random variation that determines the network patterns in predictable ways and allows for causal inference.  The challenges there are both in finding natural variation that is substantial enough to provide for a powerful experiment, and being sure that the variation affects only a limited set of variables in predictable ways, such as changing the network but not other factors determining behaviors of interest.

\subsection*{Overview}

We begin by developing some mathematical vocabulary to describe network effects, our estimands of interest (Section \ref{sec:preliminaries}). We use the language of exposure maps (and the closely related structural causal map) which assigns to every individual potential behavioral outcome of interest, $Y_i$, that depends on the
their ``treatment status,'' $D_i$, (e.g., their initial information, or some demographic variable that might influence behavior, are they vaccinated, etc.) and that of the other individuals and the network structure, $G$. The exposure maps describes the boundaries and nature of how the global vector of treatments, $D_{1:n}$ and the network $G$ maps into a smaller quantity, $d_i = f_i(D_{1:n},G)$ that maps to the outcome of interest: $Y_i(d_i)$ where $d_i \in \{1,\ldots, K\}$. The exposure map plays a central role because it provides a vocabulary to think about how much interdependence there is in the setting. For instance, if the network is such that $d_i(D_{1:n},G)$ and $d_j(D_{1:n},G)$ tend to be systematically highly correlated for most $i,j$, then there is not much separation. Treatment is essentially an aggregate shock and nothing is identifiable. This would be the case, for instance, if one had a single network and there was perfect diffusion: seeding anyone with information informs essentially everyone. Therefore, the exposure map determines whether or not one can make headway statistically in terms of studying the environment. One strand of the  literature works out related concepts (e.g., structural causal maps) so that we may cut through interdependence and estimate quantities of interest \citep{ogburn2024causal,auerbach2021local}.

We use this vocabulary  to talk about challenges for design: identification, power, data with budget constraints, collection of panels and so on. At its core, an experiment in a network setting is identified (and powered) only to the extent that one can find ``many copies'' of how a node is treated (directly and indirectly) with enough statistical independence. This requires some limits on the spillovers in the network so that the realization of outcomes has enough statistical independence (so a law of large numbers and a central limit theorem apply).
This, for instance, affects a power calculation. With no interference whatsoever we are in the usual independence case to pick the sample size $n$ ($G$ consists of all isolated nodes), if we have a clustered structure we apply the usual clustering calculations ($G$ is block-diagonal), and if we are in a mixed case the problem is more complex.
The vocabulary also sets up the researcher's choices with respect to collecting a panel rather than cross-sectional data.

Second,  we turn to the design of field experiments in Section \ref{sec:field-experiments}.
This requires considering the various types of data available to the researcher, common obstacles, and possible solutions. Given this, we look at optimal research design. We focus on concepts such as regret minimization and power, and discuss how one may try to abate the interdependence problems by constructing clusters and randomizing in intelligent ways, being cognizant of these problems \citep{ugander2013graph,viviano2020experimental,viviano2023causal,viviano2024policy,reeves2024model}.   And then we turn to the issue of measurement error. There are a variety of reasons a researcher will face measurement error: collecting network data can be expensive, there may be recall bias or survey fatigue, privacy considerations may limit responses, behaviors may be observed with noise and in ways that correlate with the network, and so on. So not only is the network potentially unknown in field work, it is also difficult to measure accurately. As research has shown, with poor measurement, statistical conclusions can be warped \citep{ChandrasekharLewis10}. We describe solutions such as EM algorithms and leveraging partial network data with statistical techniques \citep{reeves2024model}.

Third, we look at lab experiments. We explore the various kinds of questions they enable us to study given the tight control over network structure: granular social learning studies, trading on specific topologies, coordination, and so on. They also allow us to look at how meta-information structure influences behavior in a network, which is very difficult to control in the field. Further, we note that there's very little issue with power and statistical inference. We survey a collection of successful works demonstrating the types of questions they're able to answer and also pay attention to the fact that the ideas born out of the lab experiments often provide meaning for insight and fieldwork.

Finally, we briefly discuss natural experiments that provide variation in network structure that can be useful in identification.  For instance, random assignment of students to different classrooms, or refugees to cities, etc., that have different demographic compositions and hence different treatment vectors.  Such variations can provide the independent variations required for identification and statistical inference, but require enough subjects with sufficient variation and plausible independence across various individuals and their outcomes.

We end the chapter with a discussion of the value of building a corpus utilizing all three forms of experiments. By identifying key forces through lab experiments, studying how they play out through lab-in-the-field or lightweight field experiments, we are positioned to study large-scale policy implications either through natural or larger-cale policy field experiments. Without the former projects, the hypotheses necessary to pursue the latter might never emerge, especially in a network setting where there is  much to disentangle.

\section{Networks, Interference, and Identification}\label{sec:preliminaries}

\subsection{A Framework}

Each individual has some behavior or outcome of interest: e.g., a belief, how much they contribute to a public good, whether participate in some program, whether they get sick, etc.
This is denoted by $Y_i$.

They also have some treatment status, $D_i$, 
 which here is a very broad term. 

The individual’s potential behavior or outcome, $Y_i(\cdot)$, also depends on the treatment status of a set of other individuals (the vector $D_{1:n}$), the network (a graph $G$, which could be some weighted and directed graph, or a simple graph, etc.), and individual characteristics $X_{1:n}$.   Note that $i$'s outcome could depend on $X_{-i}$.

It is possible that as one changes treatment status of others and/or the network, that a person's outcome changes
\[
Y_i(D_i,D_{1:n},X,G) \neq Y_i(D_i,D_{1:n}',X',G').
\]
This violates what is known as the usual stable unit treatment value assumption (SUTVA) due to \cite{rubin1974estimating,rubin1980randomization}.  SUTVA requires that an individual's outcome only depends on their own treatment status.   Without the SUTVA assumption, outcomes are correlated and we cannot apply basic and standard statistical techniques that treat behaviors as if they are independent across individuals.

In what follows, we generally suppress dependence on $X$ unless otherwise noted, but it should be understood to be implicit.

To fix ideas about the notation, we might be interested in whether people participate in some program $Y_i$, where their decision is influenced by the decisions of their friends the $Y_j$'s, which in turn depend on the decisions of their friends and so forth through the network $G$.
Or as another example, we might be interested in whether people are eventually informed about some issue, $Y_i\in \{informed, not\}$, and how that depends on the network $G$ and who is given some initially information which is encoded in the $D_i$'s.
As yet another example,  we might be interested in how people choose a level of education $Y_i$ as their position in the network changes, or the structure of the network becomes more segregated, ...---so, $G$ is varied.

Given the violation of the SUTVA, it is not immediate that quantities of interest are identified    \citep{manski2013identification,aronow2017estimating}. Even if they are identifiable, the setting may need to satisfy a number of conditions to ensure that there is sufficient statistical power to detect spillovers of interest and test hypotheses \citep{athey2018exact,chandrasekhar2023general}. Just as in time series or spatial statistical settings, the excessive correlation intrinsic to network data effectively reduces the number of observations available to the researcher. This challenge affects data collection, power calculations, treatment allocations, and so on, as we discuss below.%
\footnote{To take things even a level further,  it can be that people's $D_i$s are co-dependent.  For example whether people become vaccinated, or attend some information meeting, etc., could be influenced by their friends.  This can be dealt with, but needs to be modeled.  For example, see \cite*{jackson2024adjusting}.}

The framework we describe above does not include all the applications that we are interested in analyzing.  For example, it can be that we are interested in how the network itself is formed:  who chooses whom as friends.  Trying to squeeze that into the notation above is a stretch, but these are still of importance, and so we discuss some of them as we proceed.

\subsection{Interference}

The goal of any experiment with network data is to estimate how some set of response variables are affected by variation in overall treatment---including network positions in $G$, changes in the network $G$, and treatment status $D_i, D_{-i}$.
The experiment can control some of the variation in this exercise, but the principal challenge is that without further conditions on the network and treatment assignments, we are unsure of what is identified and sufficiently independent for estimation.
For example, if outcomes of the nodes in a network a strongly correlated with those of their neighbors and a researcher only sees a single instance and the network is not sufficiently large compared to the number of relationships, there might be no natural way to obtain consistent estimators and conduct inference.

The intuition behind how we make progress on this is that
we need $Y_i$ to depend (``nontrivially'') only upon the treatments in some limited neighborhood of the network, and then have a large enough network so that there are sufficiently many non-overlapping neighborhoods to be able to get many observations that are roughly independent of each other.
Having the network actually be many separate networks---e.g., separate villages as in \cite{banerjee2013diffusion,banerjeecdj2019}---is one way this can work, but more generally one might simply have one sufficiently large network. If the peer influences deteriorate rapidly enough with distance in the network, and there are not too many different treatment values, then one can still estimate the treatment effects consistently.

To fix ideas,  suppose that a researcher wants to estimate peer influence in some decision, either in the field, lab, or in some natural experiment. To begin the illustration, let's first ignore the $D_i$s and suppose that the network/peer effect is what is being analyzed.
Subject $i$ is making some choice $Y_i\in \mathbb{R}_{+}$, with the vector
of actions $\mathbf{Y}$.
Suppose that subject $i$'s payoff is
\begin{equation}
Y_{i}\varepsilon_i-\frac{1}{2}Y_{i}^{2}+\beta Y_{i}\sum_{j}G_{ij}Y_{j},\label{eq:game}
\end{equation}
where $G_{ij}$ is the strength of an observed/controlled network interaction between agent $i$ and $j$, $\varepsilon_i $ is an idiosyncratic term of an individual's propensity to action,
and $\beta$ is the peer intensity.   This is a variation on the linear-quadratic network game analyzed by \cite{ballester2006s}

For $\beta$ that is not too large (in particular, is smaller than the reciprocal of the magnitude of the first eigenvalue of $G$), there is a unique equilibrium of this game.
Note that a best response (taking the first order conditions from \eqref{eq:game}) is characterized by
\begin{equation}
Y_{i}=\varepsilon_i+\beta \sum_{j}G_{ij}Y_{j}.\label{eq1}
\end{equation}
Then iteratively substituting (\ref{eq1}) we get
\begin{equation}
Y_{i}=\varepsilon_i+ \sum_{t=1}^\infty \sum_{j}\beta^t G^t_{ij} \varepsilon_j.\label{eq2}
\end{equation}

Thus, the influence of $j$'s idiosyncratic term on $i$'s action
is related to $\sum_{t=1}^\infty \beta^t G^t_{ij}$.   If $\beta$ is not too large,
and the network is not too dense so that $i$ and $j$ are far enough apart and not connected by too many walks in the network,  then $\sum_{t=1}^\infty \beta^t G^t_{ij}$ vanishes as $i$ and $j$ become further apart in the network (in a sense dictated exactly by this term) and their actions become less correlated.

Now let's return to the question of estimating treatment effects to different nodes.  Suppose that we are trying to estimate
the impact of an intervention designed to influence the $\varepsilon_i$'s.
For example, suppose that
$$\varepsilon_i = \alpha D_i + \eta_i,$$
where $\alpha$ is an effectiveness of the treatment and $\eta_i$ is noise.
This together with (\ref{eq2}) then leads to\footnote{There are many problems with trying to directly estimate (\ref{eq1}), including those outlined by \cite{manski1993,jackson2008social,bramoulle2009identification,angrist2014perils}.
The formulation (\ref{eq3}) avoids those problems to the extent that the $D_i$'s are randomized or exogenous to the rest of the setting (including the network).  In cases where the $D_i$'s are themselves endogenous, then one needs to take that into account as otherwise the estimation can be severely biased, as discussed in \cite{jackson2024adjusting}.}
\begin{equation}
Y_{i}(D, G) = \alpha D_i + \eta_i + \sum_{t=1}^\infty \sum_{j}\beta^t G^t_{ij} (\alpha D_j + \eta_j).\label{eq3}
\end{equation}

If one wants to estimate the impact of the treatments on the nodes' behaviors, one can see from (\ref{eq3}) how there is correlation in the treatment effects, but that it deteriorates with distance (for low enough $\beta$).     One needs sufficiently many observations of a given treatment that are sufficiently far apart from each other in the network that their interference becomes small enough so that one can apply an appropriate Central Limit Theorem \citep{chandrasekhar2023general}, or deduce consistency from another approach. 
The details of what is needed depend on which effects or parameters are being estimated, and what the interaction structure looks like, and so forth,
but the idea should be clear.

Note also that a treatment is no longer just the individual's treatment, but also what the treatment of their neighbors is, etc.  For example, the impact of treating a single individual is different from the impact from treating them and all their friends.  Thus, the number of ``types'' of treatments that one is interested in can, at least hypothetically, explode.  One might need to estimate a different effect for a treated node that has one treated neighbor, compared to having two treated neighbors, compared to having three treated neighbors, and so forth.  One might even need to trace out further in the neighborhoods if nodes at distance 2 interfere nontrivially, and then the number of different types of neighborhoods that need to be estimated grows, and many copies of each neighborhood that are far enough away from each other are needed.

If one does not have sufficiently many neighborhoods that represent each of the different treatment variations that might be important,
then one needs another approach.
In order to reduce the number of observations, for instance, one might be confident that the effects are captured by some parametric formulation that helps reduce the number of different configurations needed to identify the key treatments or parameters of interest.

Note that in this example, it is presumed that treatments all affect individuals via the same $\alpha$ and the network via the same $\beta$.   If these are the only two parameters that need to be estimated, then one no longer needs to treat all the individual $Y_i(D,G)$'s as the objects of interest to be directly estimated, but that one can estimate the $\alpha, \beta$, and then indirectly estimate the treatment effects.
This can be done with much less information, as different $\alpha, \beta$ pairs induce different joint distributions over the $Y$'s and one can then use observations of the resulting joint distribution to estimate the parameters.  The estimation no longer depends on a law of large numbers on individual behaviors, but on sufficient variation in the joint distribution, which can be substantially less restrictive.

This shows a trade-off:  imposing a specific model that jointly governs all behaviors can lead to parameters that can be estimated from a smaller network, despite dependencies,  but it only works for parsimonious models and the researcher needs to be confident that the model is not too misspecified.

At the heart of the above formulation, and in general in any networked analysis, is an exposure map.
An exposure map is a function that takes the treatment assignment vector and the network structure and then maps this to an exposure function.
In the above example,  (\ref{eq3}) is a sort of exposure map that provides very specific information not only about the exposures but their precise sizes.
More generally, such a map tracks how each node is exposed to the treatment in a manner that cares about both the network structure and how treatment is distributed through the network.

In summary, in order to be able to make progress in estimating treatment effects, one must either have confidence that (i) that the network effects are somewhat limited relative to the overall size of the network and that one has sufficiently many approximately independent observations of neighborhoods of the relevant treatments (which depends on the heterogeneity of the treatments, the density and neighborhood structure of the network, and the size of the subject pool),  or (ii) that the interference takes on a specific structure with unknown parameters in a way that generates nontrivial changes in the joint distribution of behaviors in a known way as one varies the parameters given the network and treatment configuration.

\subsection{Power}
A challenge stemming from network interference, as discussed above, is that it generates correlations in the outcomes in response to treatment. In the usual case, when doing power calculations, a researcher assumes a form of independence via the stable unit treatment value assumption (SUTVA)  with some posited distributional assumption.
For example, in a case in which $D_i$s take on values either 0 or 1 then one presumes that: \[
    Y_i(D_i,D_{1:n},G) = Y_i(D_i), \ \text{ with }  Y_i(1)-Y_i(0) \sim F(\tau,\sigma^2).
    \]
In terms of the exposure map, $j$'s exposure status only depends on $j$'s treatment status.
Assumptions on effect size and the dispersion allow the researcher to identify an appropriate sample size to be powered at a prescribed level for a \emph{minimal detectable effect size} \citep{duflo2007using}.

Now, let's consider the network case with interference.
Let's start by discussing the opposite extreme in which there is full interference among all nodes.   For instance, consider a treatment of informing nodes about some program and we wish to understand how being informed vs not affects behavior.  Suppose also that if a given node is seeded with information, it spreads so extensively through the network that everyone effectively becomes treated. 
Even with a large $n$, there is no way to learn how the treatment changes behavior because either everyone is treated or everyone is untreated.

Having variation in treatments that are spread through the network (and not overly interfered by communication, etc.), and with sufficiently many separated neighborhoods with limited interference across those neighborhoods, represent the key case that applies to many settings.
Namely, the are groups or clusters, with $C_i$ denoting $i$'s cluster, for which (at least approximately)
\[
    Y_i(D_i,D_{1:n},G) = Y_i(D_j: \ j \in C_i, G_{\mid C_i}).
\]
In the form written here, there is still considerable flexibility within clusters for the effects to co-depend on the network restricted to the cluster itself; clusters however are sufficiently independent.\footnote{See \cite{chandrasekharj2018} for a discussion of consistent estimation when the interference can include all nodes, but for which for each node it is sufficiently limited outside of the set $C_i$,  and the $C_i$s can overlap across nodes.}

The burden of designing a powered experiment means that one needs to take   advantage of effective near-independencies across parts the network, when available. That way the researcher has ``many measurements'' which can be aggregated to deliver consistent estimates and the ability to conduct inference. There are two ways one can do this: (1) obtain data from many, independent networks; (2) cleave a network (limited exposures) in some way that effectively generates sufficient independence. These can obviously be mixed, with case 1 being the more conservative and case 2 being the most structural, assumption-based approach.

Which strategy should be used depends on context. Lab experiments afford the ability for the researcher to construct (many) independent networks and therefore they can easily live in case 1 \citep{choi2016networks}. In the field setting, such as classrooms or villages, one may credibly be able to obtain data from many effectively independent networks. It is still tricky. Villages may be proximate and the economic phenomena (perhaps risk sharing or diffusion) may spill over, and there may be nontrivial numbers of ties across villages.
 It could also be that nearby villages are subject to some common unobserved shocks that affect outcomes.   In a given school or university different classes may not be credibly independent. Nonetheless, these kinds of settings often allow for researchers to credibly deal with the interference problem to the extent that overlapping ties or common influences are highly unlikely.

\

\subsection{Are there spillovers?}

Before delving into the specifics of field and lab experiments,  a  basic place to start is to gauge the level of externalities and spillovers. In particular, the structure of an experiment allows one to conduct exact finite sample hypothesis tests for what are called nonsharp null hypotheses.   These include null hypotheses of the form that the outcome of node $i$ is independent of the outcome of nodes at more than some distance.  The nonsharp aspect is that the null hypothesis does not imply that knowing the outcome at some treatment vector necessarily ties down what the outcome would at any other treatment vector, so it only ties down what the outcomes are to a limited extent
 (see, e.g., \cite{athey2018exact,basse2024randomization}).
An example of a sharp hypothesis is that the outcome is independent of treatment - so treatment is completely ineffective.  In that case, observing the outcome for one treatment vector implies it should be the same for any other treatment vector, under the null hypothesis.
The idea of the approach to testing such nonsharp hypotheses is to use an artificial experiment that is carefully designed to turn a nonsharp hypothesis into a sharp one.  This can be done for some nonsharp hypotheses, depending on the network and the hypothesis.
The idea is to use randomization inference with details adapted to the setting with interference.

To see why this approach is powerful in the network experimental design setting, consider the following three null hypotheses:
\begin{enumerate}
	\item No Treatment Effects:
	
	$Y_i(D_{1:n}) = Y_i(D_{1:n}')$ for all $i$ and $ D_{1:n}, D_{1:n}'$. 
	
	\

	\item No Spillovers:
	
	$Y_i(D_{1:n}) = Y_i(D_{1:n}')$ for all $i$ when $D_i = D_i'$.
	
	\
	
	\item No 2nd and Higher Order Spillovers:
	
	$Y_i(D_{1:n}) = Y_i(D_{1:n}')$ whenever $D_j = D_j'$ for all $j$ with $\text{dist}_G(i,j)<2$.
\end{enumerate}

The first hypothesis is sharp but the other two are not.

To see how the approach works, suppose that the network consists of dyads:  each person has exactly one friend, and suppose the null hypothesis is that a person's outcome does not depend on their friend's outcome - so the no spillovers hypothesis.  This is a nonsharp hypothesis, since seeing a person's outcome for one treatment does not tell the researcher what it should be for another treatment.  But it does imply that what the outcome of a given person is should be the same regardless of what the treatment of their neighbor is.  One now has essentially only half as many observations: essentially one can test whether the pairs of observations are independent.
More generally, the literature on this, then leverages that some implications of a nonsharp hypothesis to turn it into a sort of more limited but sharp hypothesis.  One loses power in terms of observations, but still has an exact implication that implies a well-defined test statistic under the null hypothesis.

In particular, observe that in each of these cases, one can in principle find other assignment vectors $D_{1:n}'\neq D_{1:n}$ that leave the outcomes for the some limited set of $i$'s fixed. For example, in case 2, no spillovers,   whatever node $j\neq i$ is assigned should not impact $i$. And in case 3, no higher order spillovers, conditional on the treatment status of one's neighborhood, it does not matter what other nodes are assigned.

The suggested approach is to create an artificial experiment in order to develop exact $p$-values for these non-sharp null hypotheses. The artificial experiment does three things:
\begin{enumerate}
    \item Pick a set of ``focal'' nodes $(V_F)$ whose outcomes are allowed to potentially change across alternative treatment vectors under the null.
    \item Partition the set of treatment vectors into equivalence classes (or strata) where $D_{1:n} \sim D_{1:n}'$ means $Y_i(D_{1:n})=Y_i(D_{1:n}')$ under the null, for all $i \in V_F$ (those that are focal).
    \item Draw new randomizations from the implied probability distribution in (2) and calculate the share of times the test statistic of interest (the researcher's choice) is as extreme or more extreme than the magnitude calculated under the synthetic draws.
\end{enumerate}

This perspective is useful both before and after the experiment, as it helps the researcher conceptualize where interference should and should not occur, and how to test for it.   And it is useful both for when the network is fixed and when the randomization is used to generate peer groups.   From an experimental design perspective, if one knows the network structure and one has some hypothesized notion of how spillovers may operate, then  this approach can be used to probe design strategies such as power calculations.  And the approach is useful for  ex-post analysis.

The previous discussion was far from non-parametric. Clearly, structure was placed both on the nature of the spillover structure and also the statistical exercise itself. But  only some types of spillovers can be tested for with sufficient power under this sort of method. For power it is important that nodes in the set $V_F$  are ``well-separated'' relative to the network effects. In practice this can mean that there may be few configurations of focal nodes that can really yield high quality results and so the appropriateness in some sense is in practice parametric.

The approach brings out a more general conceptually important exercise of figuring out what hypotheses one wants to test and what (joint) distributions are implied for at least some of the outcomes under those hypotheses.  This helps in design and power calculations.
This is where theory and experiments support each other.

As is often true in statistics, the problem is a good deal simpler if one is willing to be sufficiently parametric.   For example, if we believe that a linear quadratic game is a good description for a given setting, this provides us with a lot of structure to leverage in anticipating and calculating or simulating the distribution of outcomes under various hypotheses and designing an experiment.  Similarly, if we have a highly-specified model of diffusion or social learning, even if the network does not lend itself to easily using the tools of \citep{aronow2017estimating,athey2018exact}, the power calculation and estimation can still be accomplished.

There are two rich literatures on the econometrics of networks that we note but do not expand on beyond the fact that network experiments contribute to (and benefit from) both literatures. The first is the literature on the estimation of peer effects and social interactions \citep{manski1993,brock2001interactions,ballestercz2006,bramoulle2009identification}. The literature focuses on when, and how, one can identify spillover effects. They largely can be thought of as specific parametrizations in the exposure map or the structural causal model perspective. From the perspective of this chapter on network experiments, generating the random variation in treatment and/or network structure can provide identification in such models. For example \cite{bramoulle2009identification} identification benefits from random shocks as well as intransitive triad structure which are hard to verify in a data set, but can be manipulated with an experiment. The second is the literature on the econometrics of network formation (e.g., \cite{menzel2015strategic,leung2019,depaulart2018,sheng2020structural,depaula2020econometric,chandrasekharj2016}). Again, details of the setting have significant implications for identification, and the control of an experiment can provide conclusions that observational data often cannot.

\subsection{Data Considerations}
\subsubsection{Links and Demographics}

When studying networks, it is essential to collect rich data that capture the appropriate interactions that involve the relevant externalities and influence.

Especially in the context of field experiments, the researcher has limited control of the network. What constitutes a meaningful interaction  can be rather wide ranging. So, it is essential that the researcher be flexible and detailed in the collection of the network data. For instance, one may hear about a new job, not just through named information links, but also through kin, coworkers, or casual acquaintances \citep{granovetter1973strength}. In this kind of case, many dimensions matter for the diffusion of jobs and simply collecting ``information networks'' would be incomplete.

Generally, in the field it is impossible to fully capture the relevant interaction network.  At best a researcher is faced with working with a network that represents a portion of the relevant interactions.  Moreover, it can be that some of the captured links are actually irrelevant.  This does not mean that the researcher cannot proceed.  It simply means that the researcher faces nontrivial measurement error in the analysis.   There are ways to both impute relationships \citep{de2017community,krause2018missing,breza2020using,smith2022network} and to account for missing data \citep{chandrasekharl2010}.  Obviously, it is better to understand the biases that might be present in the analysis than to simply ignore the network and pretend it is not important.

Because the notion of a network link can vary with the application and be wide-ranging and yet it is important to be precise about the exposure map for analysis, having rich demographics can be valuable.  It allows one to construct other possible avenues for exposure; for instance, people may be likely influenced by others in their sub-caste or those in their classroom even if they're not listed as friends.  From the missing data perspective, demographic data can be used to help model missing information in the network or aid in the construction of instruments in order to identify network effects.

Third, demographic data is often useful because most socioeconomic questions involve behaviors that vary heterogeneously across demographics. Because an experiment randomizes, by construction heterogeneous effects of treatments can be identified under mild assumptions since they are exogenous to the randomization.
However, for instance, individuals that are wealthier versus poorer can be in different network positions \citep{chettyetal2022I}, as well as for many other differences in demographics.  They may also be differently influenced by some treatments.  Not tracking the demographics does not allow one to identify the heterogeneous treatment effects, and various forces end up conflated.

Fourth, collecting various demographics, especially for data sets that will eventually be made public, gives future researchers a lot of power to investigate questions that were not asked in the original investigation. In particular, when running an experiment there is often (perhaps with good reason) some restraint placed at the design stage as to which heterogeneous treatment effects are to be examined by the original researchers. But given the very complex dynamics underlying the data generating process, the returns to having rich demographic data is extremely high as it lets researchers develop future hypotheses. For instance, in \cite{banerjee2013diffusion}, nearly 17,000 households were included and while the core analysis was on the spread of microfinance through the lens of a simple SIR-style model, the richer demographic data has been used alongside the network data in numerous follow-ups. For instance, with rich data one can start to ask in this sample about social structure more generally: e.g., are wealthier less multiplexed, are balanced sub-groups more homophilous, do women build social capital differently since they marry across villages...?

Finally, people can be involved in many different types of interactions,  something that sociologists refer to as ``multiplexing''.   For instance, a household might borrow money from one household, but get information from a different one (e.g., \cite{banerjeecdj2013}).   These different layers of interaction can matter differentially in a process such as diffusion \citep{chandrasekhar2023multiplexing}.
If one only collects data about something as vague as ``friendships,'' it can mix together heterogeneous forms of interaction and may end up with biased measurement of the true interaction patterns as a result.

It can also be hard to subjects to answer an abstract question such as ``who are your close friends.''  This might also be subjectively interpreted differently by different subjects.  Asking them who they share meals with, have conversed with, have borrowed from, etc., are  more objective questions and also are cognitively easier to answer.




\subsubsection{Panel Data}

In settings in which treatment can affect the network structure,  there can be value in collecting data over time.   More generally, having multiple measurements of a network at different points in time can also help identify how things like changes in network positions change outcomes, especially in settings in which the multitude of network positions (or background demographics) are so high that cross-sectional identification is tricky.

Collecting panel data also becomes very important in situations such that the network can change between treatment and final measurement.  One can account for static heterogeneity across nodes (and groups of nodes), which may matter a lot since network formation is endogenous. However, the subtlety is that when the network changes in response to treatment, the problem is just more complicated and the research question turns into one that includes network change.

The researcher needs to ask themselves if they have the resources to collect multiple waves of data, and how does the answer change the design?  To do this, first, they have to consider what are the key variables that might be changing over time and how can those be tracked.

Another aspect of this is that the timing of interaction and activation can be important (e.g., \cite{akbarpourj2018}) and the frequency of interaction can also vary and matter (e.g., \cite{harari2016using}).

Additionally, people's networks evolve over time in ways that can be interesting.  For instance, homophily patterns in a college setting can differ depending on which year one examines, and so collecting waves of data not only helps one understand how the network evolves, but also how that affects outcomes like grade point averages \citep{jacksonnsy2022}.

\subsubsection{Outcome Choice}
The outcomes that a researcher needs to collect are generally at least partially determined by the theory they want to test.  However, it can be useful and even necessary to collect outcomes beyond the ones that are directly tied to the theory.
This can be useful as a more thorough test of a theory and of disentangling the mechanisms underlying it.  It can also be make the data useful beyond the existing study.

For instance, consider an example of social learning, where we see that when agents  receive an initial signal about the state of the world, then they are less likely to speak with a network neighbor \citep{banerjee2024less}.
There are at least two mechanisms that would predict this: (i) there is less marginal value in investing in conversations once a person has direct information; (ii) there is a signaling concern wherein those who are of low ability are more likely to need to converse to clear up their doubts about the initial information, and so people refrain from conversations to avoid being inferred to be of low ability. Both stories predict lower density of conversation, but while the first would predict weakly higher knowledge nonetheless, the second would predict lower knowledge \citep{banerjee2007public,banerjee2024less}.
If the researchers only collected information on the number of conversations (the degree in the conversation network), then it would be difficult to understand which was the driving force. But if both conversation volume and knowledge were recorded, then these two mechanisms can be distinguished.

There is also considerable value in collecting rich outcome data, particularly when the treatment potentially changes the network.  To see this consider the example of risk-sharing networks.
Consider a setting in which some villages are treated by being given access to formal financial markets, and other control villages are not, and a researcher is interested in understanding whether access to formal markets changes informal networks \citep{banerjee2024changes}.    If the researcher sees different patterns of networks in the treatment and control villages, then by also seeing changes in patterns of people's consumption variance (beyond those who access the financial markets) then researcher can infer that network changes have welfare consequences. A different interpretation of this is that it is often useful to not simply think of network data as the only outcome of interest even when it is being measured in reaction to some treatment, as it can be co-determined with other outcomes.

More generally, planning to make data public can affect the design.
For example 
\cite{chettyetal2022I} presents calculations of some quantities 
using measures that were coarser than what we might have otherwise used in order to release the data without running into privacy concerns.   Anticipating that one would like to release the data, can affect which data sources are used to couple with an experiment, or which field sites are chosen, etc.
The choice of population can also be made to enrich the existing data around the world, and to complement previous research.

\subsection{Factorialization}

Networks are famous for exhibiting complex changes in outcomes.
Examples include: phase transitions whereby slight changes in network density result in dramatically different behaviors and outcomes; different diffusion dynamics based on the complexity of the contagion process; co-evolution of common knowledge and diffusion of information; sensitivities in matching and stability concepts, among other things.

For example, as one changes who or how many people are seeded in a social learning process, the level of common knowledge changes thereby affecting the downstream behavior in non-obvious ways \citep{awaya2024spreading}. To take another example, cooperation across social groups (e.g., race, ethnicity) can be non-monotone in the population share. When the population share is very low or high, there is less insularity, and in contrast homophily and network divisions between groups may be maximized when different groups are of more equal sizes \citep{currarinijp2009}.
Diffusion of information about upcoming vaccination camps may only be successful when paired with reminders as well: so that the network behavior interacts with the reminders, with neither
being successful alone \citep{banerjee2021selecting}.

The general point is that because networked interactions can depend on many complicated features,
it can be necessary to run an experiment that is ``high dimensional'' in terms of both treatment and the data collected.
Without doing so, we might not be able to identify distinguishing features of one theory versus another, and it can also be that only specific {\sl interactions} of variables lead to treatment effects.

To see this, consider an example in which the researchers
are trying to increase participation in a program.   They have the option to vary (i) what share of people are ``seeded'' in a network in terms of being given initial information (1-3 with 1 being a low fraction and 3 being a high fraction); (ii) how intensively the program is advertised to the general population (1-3, again 1 being low);
(iii) how detailed the information about the program that is conveyed is (1 low, 3 high).
So in principle there are 27 different cells plus a pure control, so 28 different possible combinations to be tested. The experiment is at the network-level, so following a naive rule of thumb, such as 30 independent observations per combination to be tested, the researcher would anticipate needing 840 independent networks.

The natural impulse is to avoid this and pare down the set of treatments and heterogeneities to be explored for the sake of power. The act of doing this, however, is equivalent to saying that we do not find it worthwhile to learn about the treatment cells that we drop. If out of 27 treatments we simply conduct a $2 \times 2$ experiment, we are opting not to learn about the 23 other treatments. Either this is then only one part of a larger experiment and cannot draw conclusions, or the researcher must be confident that they already know which of the 27 do not require any study. That may be the case sometimes, but especially in network settings where there are externalities and natural complex interactions, we often are unable to pre-determine which seeding/ advertising / complexity combinations yield better or worse outcomes.

A middle ground between treating every possible combination as completely independent, and ignoring a bunch of them,  is to make use of some knowledge of how combinations and dosages might work.  If there are monotonicities in how treatment effects work, and a potential that some differences in dosages or combinations might not matter,  then that actually adds a lot of structure and one can design and perform an experiment with many fewer observations.
Let us explain.

An approach is to discover which treatment bundles  yield similar effects, and thereby reduce the dimensionality of the analysis.

\begin{figure}
     \begin{tikzpicture}[scale = 0.6] \def \n {5} \def \radius {2cm} \def \margin {8}
     \node[draw, circle,  minimum size=19pt] at (0,-4) (v1){$[1,1]$};
     \node[draw, circle, minimum size=19pt] at (-2,-2) (v2){$[1,2]$};
     \node[draw, circle,  minimum size=19pt] at (-4,0) (v3){$[1,3]$};
     \node[draw, circle,  minimum size=19pt] at (2,-2) (v4){$[2,1]$};
     \node[draw, circle,  minimum size=19pt] at (4,0) (v5){$[3,1]$};
     \node[draw, circle,  minimum size=19pt] at (0,0) (v6){$[2,2]$};
     \node[draw, circle,  minimum size=19pt] at (-2,2) (v7){$[2,3]$};
      \node[draw, circle,  minimum size=19pt] at (2,2) (v8){$[3,2]$};
       \node[draw, circle,  minimum size=19pt] at (0,4) (v9){$[3,3]$};


      \draw[rotate around={45:(2,0)}, blue] (2,0) ellipse (3cm and 3cm);
      \draw[rotate around={45:(-1,-3)}, blue] (-1,-3) ellipse (1.5cm and 3cm);
      \draw[rotate =0, blue] (-4,0) ellipse (1.2cm and 1.2cm);
      \draw[rotate around={135:(-1,3)}, blue] (-1,3) ellipse (1.5cm and 3cm);

     \draw[line width = 0.3mm, >=latex] (v1) to (v2);
     \draw[line width = 0.3mm,  >=latex] (v1) to (v4);
     \draw[line width = 0.3mm,  >=latex] (v2) to (v3);
     \draw[line width = 0.3mm,  >=latex] (v2) to (v6);
     \draw[line width = 0.3mm,  >=latex] (v4) to (v6);
     \draw[line width = 0.3mm, >=latex] (v2) to (v6);
     \draw[line width = 0.3mm, >=latex] (v3) to (v7);
     \draw[line width = 0.3mm, >=latex] (v4) to (v5);
       \draw[line width = 0.3mm, >=latex] (v6) to (v7);
      \draw[line width = 0.3mm, >=latex] (v6) to (v8);
      \draw[line width = 0.3mm, >=latex] (v5) to (v8);
     \draw[line width = 0.3mm, >=latex] (v7) to (v9);
      \draw[line width = 0.3mm, >=latex] (v8) to (v9);

\draw[ rotate = {45}, purple, line width = 0.75mm] (-1.15,1.6) rectangle  (4.15,4);

    \end{tikzpicture}
     \begin{tikzpicture}[scale = 0.6] \def \n {5} \def \radius {2cm} \def \margin {8}
     \node[draw, circle,  minimum size=19pt] at (0,-4) (v1){$[1,1]$};
     \node[draw, circle, minimum size=19pt] at (-2,-2) (v2){$[1,2]$};
     \node[draw, circle,  minimum size=19pt] at (-4,0) (v3){$[1,3]$};
     \node[draw, circle,  minimum size=19pt] at (2,-2) (v4){$[2,1]$};
     \node[draw, circle,  minimum size=19pt] at (4,0) (v5){$[3,1]$};
     \node[draw, circle,  minimum size=19pt] at (0,0) (v6){$[2,2]$};
     \node[draw, circle,  minimum size=19pt] at (-2,2) (v7){$[2,3]$};
      \node[draw, circle,  minimum size=19pt] at (2,2) (v8){$[3,2]$};
       \node[draw, circle,  minimum size=19pt] at (0,4) (v9){$[3,3]$};


      \draw[rotate around={45:(-1,-1)}, red] (-1,-1) ellipse (3cm and 5cm);
      \draw[rotate around={45:(2,2)}, red] (2,2) ellipse (1.5cm and 4.5cm);

     \draw[line width = 0.3mm, >=latex] (v1) to (v2);
     \draw[line width = 0.3mm,  >=latex] (v1) to (v4);
     \draw[line width = 0.3mm,  >=latex] (v2) to (v3);
     \draw[line width = 0.3mm,  >=latex] (v2) to (v6);
     \draw[line width = 0.3mm,  >=latex] (v4) to (v6);
     \draw[line width = 0.3mm, >=latex] (v2) to (v6);
     \draw[line width = 0.3mm, >=latex] (v3) to (v7);
     \draw[line width = 0.3mm, >=latex] (v4) to (v5);
       \draw[line width = 0.3mm, >=latex] (v6) to (v7);
      \draw[line width = 0.3mm, >=latex] (v6) to (v8);
      \draw[line width = 0.3mm, >=latex] (v5) to (v8);
     \draw[line width = 0.3mm, >=latex] (v7) to (v9);
      \draw[line width = 0.3mm, >=latex] (v8) to (v9);

      \draw[ rotate = {45}, purple, line width = 0.75mm] (-1.15,1.6) rectangle  (4.15,4);

    \end{tikzpicture}
    \caption{\textcolor{blue}{$\E[Y_i \mid (2,3)] =\E[Y_i \mid (3,3)]$} (Left) and  \textcolor{red}{$\E[Y_i \mid (2,3)] \neq \E[Y_i \mid (3,3)]$} (Right) \label{fig:factorial}}
\end{figure}
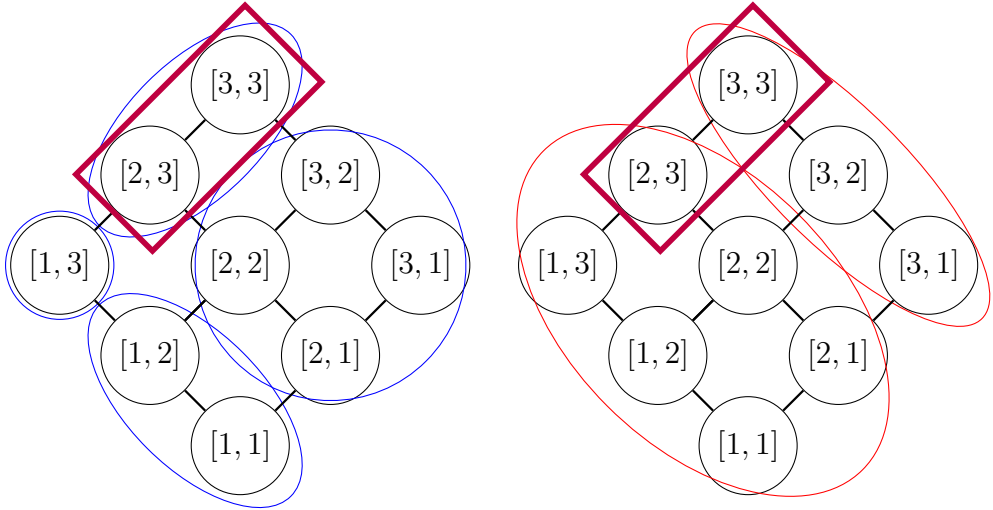

This corresponds to a {partition} of the space of heterogeneity induced by the experiment.  Figure \ref{fig:factorial} presents this graphically for the case of two arms each varying from intensities 1-3, omitting the third arm.
For instance, imagine that if there is a larger impact if a high amount of nodes are seeded, but that the a low or middle amount lead to the same outcomes.
Also imagine that a medium and high amount of advertizing have the same impact, but that they always have a higher impact than a low amount.
There could also be an interaction effect where the higher levels of dosage of both have a much greater impact than the sum of the effects.

This is pictured in the left-hand panel of Figure Figure \ref{fig:factorial}.
The nodes represent the various treatment bundles and nodes that are in ovals together are ``pooled''---they have the same expected outcome.

A different configuration is the one on the right.  Here the level of advertising never matters, but the level of seeding does, once it hits the third level.
On the left we see that (2,3) and (3,3) can be viewed as the same treatment and on the right we can see that, instead, (2,3) and (3,3) are \emph{different}. Notice that very different underlying theories would rationalize such different patterns of heterogeneity, and therefore partitions. But one cannot ever study this in the first place without running factorial design.

Importantly, one can use Lasso, Rashomon Partitions, and other techniques to determine which treatment bundles should be pooled, under some basic monotonicity assumptions.  This can greatly reduce the number of observations needed.
This is the subject of \cite{banerjee2021selecting} and \cite{venkateswaran2024robustly}.

In settings such as experiments on networks, where the heterogeneity space can be so complex, one needs to be able to run more intricate experiments and regularize in a data-driven way rather than forcing regularization ex-ante by ignoring important treatment combinations.

\subsection{Networks themselves as Treatment or Outcome}

As we have mentioned in a few of our examples so far,
there are settings in which the goal of the experiment
is either to see how different network structures affect behavior,  or
that we are interested in
seeing how the network itself is formed or reacts to some treatment.

Let us first discuss how networks are affected by treatment, and how that influences experimental design.

\cite{comola2021treatment} and \cite{banerjee2024changes} provide two examples in which the interest is in how a treatment (access to savings and microfinance, respectively) affects the networks themselves, as well as resulting welfare.

The settings are different in a number of ways.
A key distinction is the number of networks in each study. In \cite{comola2021treatment}, the authors study data from 19 villages in rural Nepal, whereas in \cite{banerjee2024changes} the authors study 75 villages in Karnataka and 102 neighborhoods in Hyderabad. This motivates different strategies by the authors.

\cite{banerjee2024changes} analyze two settings: 75 villages from Karnataka in which microfinance entered over half of the villages and 102 neighborhoods in Hyderabad where microfinance was randomized to half of the neighborhoods. The Karnataka sample contains both a before an after snapshot of the network,  so direct measurments of how the networks change whereas the Hyderabad sample consists a single time view of the network, but then compares control with treated.  This allows one to test and reject various theories about network formation, and to develop a new model to explain what is observed.
Given the high number of villages and the randomized treatment in the Hyderabad sample, one can test several hypotheses and do it without imposing paramateric assumptions. This is both a luxury afforded by the sample size.

\cite{comola2021treatment} are working with a smaller number of villages and so instead explicitly model interactions a linear-in-means network model \citep{bramoulle2009identification} and evolution is modeled as well through a conditional edge-independent model with pairwise observable controls. The goal is to obtain a causal estimate of the peer effects, split by old and new links.

Both methods allow one to estimate models of evolving networks and to test various hypotheses using combinations of cross-sectional and panel observations of networks.

In the section on lab experiments, we discuss the testing of models of network formation,  as well as how variation in networks affect outcomes, in more detail.



\section{Field Experiments}\label{sec:field-experiments}

When designing field experiments with networks, there are a number of factors that the researcher has to consider. (These considerations supplement the discussion in  Chapter 1 on general experimental design.)

First, and most importantly, the researcher needs to decide whether they are to be in a situation with many independent networks or a single large network.
Conditions that need to be satisfied regarding the nature of the network effects, both in terms of identifiability and also statistical power, differ across the two regimes.
In broad strokes, it is preferable to have many independent networks. Given the context, however, it might be infeasible either due to cost considerations or because the object being studied, such as in the case of social media, is in fact intrinsically  one large network.

Second, the researcher needs to decide whether they will be looking at a cross-section or a panel. There are trade-offs. Of course in the context of a panel one can better deal with node-level fixed unobservables, but on the other hand, by the very definition of a panel, one needs to understand how the network itself changes over the course of the data collection horizon.  Further, the panel may be in $Y$, in treatments $D$, or in $G$, or in in some combination. 
Each of these cases merits its own strategy.

Third, there are design considerations that depend on the details of the data itself as well as the data generating process. That is, the structure of the network driving the spillovers and the patterns of externalities and of  spillovers affect power and identification.

\subsection{Many Networks versus Single Large Network}

The most obvious factors that determine whether the researcher works in the many or single network regime are usually natural constraints of the setting, and cost of the project. For example, the research might be conducted on one social media platform that has one giant component connecting all the nodes, so by definition one is in the single large network case. Similarly, there is only effectively only one network of banks in the US.  In the latter case it could be hard to find different parts of the network that are effectively independent, and so the network has to be treated as one large correlated network observation.  In the case of social media, there can be subcommunities that are effectively isolated from each other and so they can be treated as many independent networks.

In terms of costs, there are not only costs of accessing many different sites or settings, but there can also be issues of obtaining permissions.

If the researcher has discretion in how many networks to observe, then the nature of interference, and our discussion above about whether there are enough quasi-independent observations available in one large network becomes germane, as well as how parametric and specific the hypotheses are that are being studied, and whether good identification is possible even in the face of nontrivial correlation in outcomes.

Second, the researcher should reflect on how much measurement error they are likely to face. Specifically, the researcher needs to be able to articulate how precisely they can define a link, and how accurately they can measure that link they have defined. So, for example, when studying information about a job, how precisely can they find and measure all of the edges between nodes on which information about a job can travel? Measurement error in both the definition, as well as the actual elicitation, introduce noise and bias, and push in the direction of needing more observations.  In particular, if the network is mismeasured, then two parts of the network that are thought to be independent could be more closely tied to each other.
Overall, relying on observation of a single large network works best when
the researcher is confident that the spillovers are limited and are well identified and measured in the data so that there is no slippage; or else the hypothesis is sufficiently specific so that it can be well-identified from a joint distribution of outcomes.

\subsection{Link Elicitation}

Before considering details of link elicitation, it is helpful for a researcher to first reflect on the kinds of network data that are essential for their purposes. First, does the researcher require extremely granular and extensive link data or is it sufficient to have some network-level patterns or some basic moments of individuals' neighborhood structures?  Second, does the researcher face any constraints with regards to a definition of a link? For example, links in social media or financial transaction networks are well defined: the measured links are exactly what they are supposed to be. However, if we are considering peer influence in a classroom or how information about a job travels through a social network, the measured links are much broader. One can learn about a job from one's friend or kin or RoSCA (rotating savings credit association) member or simply in the town square. Similarly, peer effects in a classroom can be mediated through channels beyond one's immediate friends that the respondent listed in a survey. Third, does the researcher face implementation constraints? For example, perhaps the researcher needs to maintain privacy, obtains masked or aggregated data, has a limited budget and therefore can only sample a small share of nodes, or has a long survey instrument that induces survey fatigue. The researcher obtains noisy (partial, aggregated, mismeasured) network data in each of these cases.

The type of network data constrains the nature of the experiment that can be implemented.

A first type of network data are those elicited from surveys. This consists of going to community such as a school, village, town, firm, and asking a sample of individuals about the nature of their interactions with others. Here, researchers can vary their strategy, ranging from asking about all of their links within the pre-specified population, just the links among the sample set to some top-coded number of links. Surveys, of course,  lead to survey fatigue, subject heterogeneity in interpretaiton of questions, recall bias, among other things, and therefore come with measurement error. However, surveys have the virtue of also being able to tailor the questions to focus on exactly the interactions of interest and therefore, most likely actual channels for exposure for a given social economic phenomenon.

The second type of data  is some form of administrative or data from media, including platforms. Here, researchers look at an organization's hierarchy or communication patterns,  or social media links, such as followers or friends. This sort of data has the upside of making it more straightforward to  obtain and define a link. The danger, however, is that the links might not include all of the relevant interactions, and might include spurious ones. For instance, there may be many routes by which one is influenced beyond who they are explicitly linked to via some particular media interface, and this can lead to omitted variable biases that correspond to errors in the exposure definition.

The third type of data is more aggregative. This could be due to privacy reasons or costs, but it is essentially another form of looking at the volume of links or interactions that are given subject has with other types of alters. This could be, for example, the volume of outgoing flows from a bank to other banks, it could be the number of interactions that a villager has with members of each of several other castes, and so on.
This might be the easiest sort of data to collect in many cases, and gives information on each subject's local network, which can suffice in some studies.  It does not permit a precise reconstruction of the overall network, and so situations in which cross-correlations need to be identified and controlled for it may not suffice.

\

Let us now discuss survey elicitation in more detail. 
We suggest that the researcher focus on the following six elements.

First, they should do a census so they are aware of the full set of people who could be interacting and affecting outcomes, and when conducting the census, they should include features/ characteristics of the respondents that  triangulate the node. What we mean by this is that sometimes when conducting surveys, basic identifiers can have measurement error. Sometimes certain nodes know others by different names, nicknames, or just other identifiers, and in some cultures many people share the same name.  By having a richer set of characteristics associated with each node in the census, one can uniquely identify any given node because it is unlikely that another has the same or a very similar vector of features. Obviously, triangulation gets easier the longer the feature vector is.

Second, the researcher needs to decide how the respondent should respond to link elicitation questions. There are several options: (1) show the entire census to the respondent; (2) have them think in general terms and provide a list of alters and then go layer-by-layer to identify which people share what kind of link with the respondent; (3) prompt the respondent layer-by-layer (with the possibility of some drop down menu, or tablet that links to the census, making identification of alters easier). This is not an exhaustive list but spans three  common techniques. The first has the downside that it can be overwhelming. The second and third trade-off survey fatigue and recall bias. There is no one-size-fits-all solution here, though in our experience (2) and (3) yield similar results whereas (1) can be difficult to execute in all but the smallest settings. In each of these cases, it is helpful to collect the respondent's responses about their alters' traits since that can help in matching if the data has considerable error.

Third, as discussed above, the researcher needs to determine how many dimensions or layers of links they want to elicit. Obviously, more granular information is  better, and this is the argument for collecting detailed multiplexing data.  But it can also increase survey length and fatigue.   There's also the benefit that if one asks a respondent about different layers than one is more likely to pick up people that the respondent is linked to but were forgotten on one interaction, but get primed when they came across a new type of interaction.

Fourth, the researcher must develop, when applicable, a word bank or script for link elicitation that yields reasonable responses.  Figuring out how to talk to the potential respondents  in order to get consistent, replicable, quantitative data, is essential  (see also Chapter 11).
If one is using multiple teams to conduct surveys, it becomes essential that they all follow the same protocol and script.  Otherwise, it could be that one team spends more time with subjects than another team, and then generates systematically higher numbers of connections, which could result in some systematic bias if teams are not randomly assigned across nodes of the network.

Fifth, in many contexts such as sharing risk it is not clear if the researcher wants hypothetical or past-realized links.  For instance, does the researcher ask ``to whom would you go to borrow X in an emergency?''  or do they ask ``from whom have you borrowed X in the last month?''  The second has the advantage of being more objective and factual, while the first is abstract and hypothetical.  However, the second may miss relationships that are important but were simply not recently active (e.g., \cite{granovetter1973strength}).    The researcher must be clear on their goal here. In the example of risk sharing, for instance, one may link households $i$ and $j$ if either has made a transfer to another in response to a recent past shock. However, this may not be sufficient since there is a different between a realized transfer and a transfer that \emph{could have} been realized. One can imagine that $i,j,k$ are mutually linked but over the duration covered by the survey shocks were such that only an $ij$ transfer was observed. It would be a mistake in this kind of environment to not consider $ik,jk$ as linked.

Sixth, the researcher needs to keep some link elicitation hygiene in mind. A common practice is to top code the link elicitation; e.g., asking a respondent to name up to five friends, which helps in both limiting the scale of collection and survey fatigue.  The problem with this approach is that if the cap has any meaningful bite, e.g., $\mathbb{E}(\text{degree}) = 4.7$ but a cap of 5, then the researcher loses considerable information about the actual network structure and has to deal with censored data bias. Like top-coding, handling the censoring ex-post is difficult and one is better off avoiding the issue through thorough piloting ex-ante.
Another issue the researcher may face is how to think about directionality. When $i$ lists $j$ but not the other way around, is this truly a directed relationship or simply appears that way because $j$ forgot to mention $i$, or has them as ranked 6th and can only name five. A simple diagnostic is to calculate the rate of non-reciprocation that exists even in obviously reciprocal links such as non-nuclear family kin.  One can also use the data to see how predictive directionality is of interaction and externalities.

A final point on hygiene: often networks are defined at levels such as the household though data is obtained from individuals.  Knowing at what level the analysis will be conducted (e.g., household or individual) and how households interact internally, and how many people within a household need to be surveyed to provide useful data, should be thought through at a piloting stage.
The researcher should also be sure that in eagerness to implement a  network elicitation survey, the they do not omit other measures of social interaction such as schools, classes, kin, RoSCAs, self help groups (SHGs), microfinance groups, and so on.

Sanity checks are essential, and pilots are extremely useful in seeing what issues arise when trying to make sense of the data.  Piloting becomes somewhat trickier in the face of preregistration, but one can still generate simulated data to see what problems might have been overlooked before it is too late.

\subsection{Partial Link Data}
In many cases, researchers end up with partial network data. The first reason may be due to cost. Censuses can be expensive, and especially so for graduate students. This may unnecessarily force an assumption of limited spillovers, while the size of a spillover ought to be a   (social) science question and not a budgetary one.  The second reason may be privacy. Policymakers, partner institutions, and even respondents may be hesitant to obtain overly granular interaction data.  To deal with these issues, a researcher could instead obtain partial data across many independent networks rather than spending that budget attempting to obtain pristine data about a single large network.  (This dovetails both with the discussion on survey design (Chapter 7) and running experiments in developing countries (Chapter 6), where survey methods are paramount.)

There are numerous approaches to sampling to get partial data. We describe three strategies here.

The first is star / subgraph sampling. For each $i$ in the sample,  star sampling elicits the entire $i$th row,  $G_{i,\cdot}$. (Induced) subgraph sampling is $G_{ij}$ where both $i$ and $j$ are in the sample.

The second is respondent driven sampling (RDS). The basic idea is to start with some node $i_0$, elicit its neighborhood, pick a random $i_1$ in $i_0$'s neighborhood, and repeat the process until a sample $i_0,\ldots,i_m$ has been drawn (along with neighborhoods for each). RDS is modeled as a Markov process, and this motivates a certain class of (reweighting) estimators.
There is a caution about respondent driven sampling.  It can lead to biased observation of nodes.  For instance, people who have more friends are proportionally more likely to be found via such a search, which induces a biased degree distribution (e.g., see the discussion in \cite{jackson2008social,jackson2019}).  Because of this, it is important to be careful and understand whether one is in a setting where appropriate reweightings (e.g., Horvitz-Thompson, \cite{rohe-rds}) can be leveraged. It can also be that there are other sorts of selection, so that people are more likely to nominate friends with certain attributes which complicate conventional reweighting strategies.

The third strategy is Aggregated Relational Data (ARD). In this approach, a respondent $i$ is asked about how many of their friends ($G_{ij}=1$) have a certain trait $k$ ($X_{jk}=1$). For example, ``How many of your friends have a motorcycle?'' and what is recorded is the count rather than the names of the friends. The outcome is, for all $i$ in the sample,
\[
M_{ik} := \sum_j G_{ij} \cdot {\bf 1}\{X_{jk}=1\}, \ X_{jk} \in \{0,1\}.
\]
This has the virtue of being extremely cheap. In calculations done by J-PAL South Asia described in \cite{breza2020using}, ARD would have allowed the researchers to save 80\% of their budget and yet obtain the same empirical conclusions in the experiments where they had full network data.  The restriction is that this technique only works when this sort of information allows for enough of the network to be reconstructed to identify the key behavioral implications.  If detailed network structure is needed for identifying behavior then more information might be needed.  Ultimately, one has to carefully consider what the minimum information needed is.

\subsection{Many Independent Networks}
If the researcher has data from many independent networks, then they face a very permissive spillover environment. Spillovers can be large and, further, fairly arbitrary in shape: $Y_{1:n}(D_{1:n})$ is ``reasonably'' unrestricted. This means that one can allow for large spillovers and correlations, allow allow for non-monotone effects in exposure to treatment, and obtain treatment effects for a variety of different situations.

Turning to the statistical problem, inference is straightforward as fundamentally the researcher just faces a standard cluster (here graph) level design. There is some subtlety here, which we return to below, due to the uncertainty induced by  partial network data. Having limited data per network could imply that the researcher needs to estimate distributions over $G$ in order to compute distributions over exposure. The quality of this exercise depends on how large  a sample was obtained per independent network and also requires that the underlying networks are dense enough for these techniques to be applicable.

\subsection{A Single, Large Network}

Turning to the case in which a researcher has an observation of a single, large network, we focus on four distinct (non-exhaustive) scenarios.
First, the researcher may know the social network and runs a single wave experiment. Second, the researcher knows the social network, and collects multiple waves of data. Third, the researcher has access to partial network information in a single wave experiment.
Fourth, the researcher does not know the network structure but is interested in a single or multi-wave experiment.

\subsubsection{Known $G$, Single wave}
When a researcher has a single large network and an opportunity to run a single wave experiment, a natural question to ask is whether they should do a cluster design.  That involves treating multiple individuals adjacent to each other in various clusters of the network, and these need to be chosen so that they do not to spillover across different clusters in the network.  Whether a cluster design is preferable to an individual treatment design is the subject of \cite{viviano2023causal}.

The experimentalist has a trade-off. On the one hand, if there are essentially no spillovers, then there is no reason to cluster the design. Potential outcomes are virtually at the individual level and SUTVA is all but satisfied. In contrast, if there are nontrivial spillovers, then clustering is useful because it reduces bias and allows the spillover within the cluster to be unrestricted or at least less restricted and then identified by looking at a large number of clusters that involve different patterns of treatments and spillovers. For that to work, one requires an assumption of a lack of spillovers {\sl across} clusters.

\cite{viviano2023causal} focus on a worst case bias-variance trade-off to help the experimentalist evaluate when to cluster rather than simply use Bernoulli randomization. They show that the worst-case bias relates to the average share of nodes' links that are across clusters. They also show that the worst-case variance relates to the inhomogeneity in cluster size: when there is considerable variation in cluster size then there is more worst-case variance. Having just a few large clusters obviously can lead to a lot of variance and low power, though it does well on bias; while conversely a Bernoulli design mechanically reduces variance but is vulnerable to (overwhelming) bias.  The authors work out specific rates and practical recommendations for experimental design.
Informally, they identify rates that give the following guidance:
\begin{enumerate}
    \item Small spillovers, few clusters: Bernoulli design
    \begin{itemize}
        \item maximum spillover rate that can be sustained vanishes relative to the number of clusters,
        \item more clusters requires smaller spillovers and vice versa.
    \end{itemize}
    \item Moderate spillovers, many clusters: Cluster design
        \begin{itemize}
        \item clusters can have order constant nodes,
        \item spillovers still need to be small, but can be non-vanishing relative to cluster size.
    \end{itemize}
\end{enumerate}
The practical implementation of \cite{viviano2023causal}'s Causal Clustering is . Given an adjacency matrix $G$, one can implement their variation of a convex relaxation for a  min-cut program. The algorithm is fast and out-performs many other ad hoc approaches in the literature.

\subsubsection{Known $G$, two-wave}
Sometimes there is a chance to experiment in a more adaptive way. This could be because there is an opportunity through a partner to run a second wave or because a pilot was larger-than-intended.  Or it could be because one has the opportunity to experiment adaptively in the setting. \cite{viviano2020experimental} looks at adaptive experimentation, specifically a two-wave experiment, in the presence of network spillovers. He explicitly considers the trade-offs between a pilot wave and a main experiment wave, offering researchers a framework to optimize their experimental designs.

At baseline the researcher observes $G$ and a set of features $X_{1:n}$. The goal then is to allocate individuals to a pilot (Wave 1), designated by $P_i=1$, select a sample for the main experiment (Wave 2), designated by $R_i=1$, and assign binary treatment $D_i$.

The procedure is roughly:
\begin{enumerate}
    \item Select individuals to the Wave 1 (pilot), $P_i=1$, such that all such $i$ have
    \begin{itemize}
        \item few connections to others, and
        \item neighbors also in pilot (for at least some).
    \end{itemize}
    \item Select individuals into Wave 2 (experiment), $R_i=1$, such that
    \begin{itemize}
        \item they were not in the pilot ($R_i=1$ means $P_i=0$), and
        \item their neighbors were not in the pilot ($R_i=1$ means $G_{ik}=0$ for every $k: \ P_k=1$).
    \end{itemize}
    \item Collect data
    \[
    R_i\cdot(Y_i,X_i,D_i,D_{j: \ G_{ij}=1},G_{i,\cdot}).
    \]
    which is a list of outcomes, features, treatment assignments, neighborhood treatment assignments, and row in the network for each individual in the experiment's sample.
    \item Build weights and construct estimand.
\end{enumerate}

The researcher needs to take into account the following considerations given a budget:
\begin{enumerate}
    \item Sample size allocation: How to divide the total sample between the pilot and main experiment.
    \item Treatment assignment in the pilot: Which units to treat in the initial wave.
    \item Data collection strategy: What network information to gather and how extensively.
    \item Treatment assignment in the main experiment: How to leverage pilot data for optimal assignment.
\end{enumerate}

Allocating more resources to the pilot allows for better network data collection and more precise estimation of spillover effects. This, in turn, enables more efficient treatment assignment in the main experiment. However, it comes at the cost of a smaller sample in the main experiment, potentially reducing overall precision.

The paper shows: (1) the larger the size of the pilot experiment, the smaller the estimation
error; (2) the larger the size of the pilot, the stronger the constraints imposed in the
optimization algorithm, and therefore the larger the regret with respect to an ``oracle''
assignment mechanism.

Interestingly, the author shows that in some cases, it may be optimal to assign treatment to units that are not the most central in the network during the pilot phase. This counterintuitive result stems from the need to gather information about various parts of the network, not just the most connected components.

The framework also addresses how researchers should adjust their strategies based on their prior beliefs about network effects. When spillovers are expected to be large, the optimal design tends to allocate more resources to the pilot to better understand these effects, which can then be used to better design the main experiment.

More generally, one might have other reasons for collecting multiple waves of data, as for instance discussed above in the context of estimating network evolution in response to treatment.  It could also be that treatment effects are dynamic, and the researcher is interested in those dynamics.

\subsubsection{Partial $G$, single wave}
Another situation a researcher may be in is where they have limited or partial data on $G$ and only a single shot to run an experiment. This could be because they have a limited budget, need to aggregate or mask data to preserve privacy, or for any number of other reasons. The main question a researcher must ask is whether with such partial data thy  are still able to design experiments to effectively estimate the key estimands of interest.

This is the subject of  \cite{reeves2024model}. The authors consider the case where an experimentalist has access to a single network and has collected partial network data (such as ARD). The estimand is a parameter vector in a structural causal model, so a generalized peer effect parameter, and the goal is to do this through OLS or GMM more generally. The framework developed allows an experimentalist to assign treatments optimally in a network, using only the partial data. Here the goal could be things like variance minimization or optimal seeding in a diffusion experiment.

The procedure is as follows.
\begin{enumerate}
    \item Use the partial data to invert parameters of a random graph model.
\begin{itemize}
    \item The paper studies a possibly misspecified stochastic block model, though the argument could be made more generally.
\end{itemize}
    \item Given parameters of a generative network model, construct expected moments in order to estimate the parameter of interest from the structural causal model.
\end{enumerate}
This approach allows the study for larger spillovers and more layered spillovers than the usual approach to exposure maps. The gains are parametric since the causal structure is being modeled, which is in itself a limitation.
However, it is useful because in some cases without imposing such a structure, an exposure map approach with such limited network data may preclude analysis of even basic spillover structures such as non-trivial contagion.

For an experimentalist who is interested in trying to decide how to allocate treatment, the authors suggest first estimating the generative network model from the baseline and then reducing the treatment allocation problem (which would be NP-hard) to an easier problem. Rather than worrying about which of the $n$ nodes should receive what treatment status, since each of the $n$ nodes are allocated to one of $J$ clusters (e.g., in a stochastic block model), then the question has become how much of the treatment budget to allocate to each of the $J$ clusters.

That is, if   we care about minimizing asymptotic variance of some estimator of  $\omega'\beta$, where $\omega$ are some weights and $\beta$ is the structural causal model's parameter of interest,
    \[
    \min_{D_{1:n}} \text{var}(\omega' \hat \beta \mid D_{1:n}, \text{Graph model}),
    \]
then we have an NP-hard problem.  But if we transform this problem into one where we have $J = o(n)$ groups and had levels $\tau \in [0,1]^J$, we are looking at
      \[
    \min_{\tau} \mathbb{E}_{D_{1:n}\sim P_\tau}[\text{var}(\omega' \hat \beta \mid D_{1:n}, \text{Graph model})]
    \]
    where $P_\tau$ is the distribution of treatment assignments under saturation vector $\tau$. The problem is no longer NP-hard, though it may run into some non-convexities that can be taken care of through Bayesian optimization procedures. The object does not have to be variance minimization either; the authors provide an additional example of experimental design with the aim of maximizing diffusion.

To illustrate the value for an experimentalist,  \cite{reeves2024model} use the network data from \cite{banerjeecdj2019} to generate synthetic outcomes and study the properties of estimators using full and partial network data under uniform randomization and using the optimal design for treatment assignment (here, seeding). The estimand in this exercise is a parameter in a diffusion/hearing model (the diffusion centrality passing parameter). The exercise shows that with full data, the variance in the estimation is considerably lower under the optimal design as compared to uniform randomization. This gap increases steeply in sample size over the support. What is striking is that not only is the gain substantial with the partial data (here, ARD), but moreover ARD with optimal design is closer to full data with optimal design than it is to full data with uniform randomization. That is, the experimentalist is better off with partial data but being smart about how to assign treatment rather than having pristine network data but assigning treatment sub-optimally (at random, which is considered standard).

\subsubsection{Unknown $G$, single or multi-wave, with few clusters}
So far we have discussed cases in which the researcher either fully observes the network or observes enough to readily construct a distribution over exposures  that is sufficient to recover parameters of interest.

A particularly interesting case, and one that is policy relevant, is in which  the network itself is unknown. In many settings
a researcher knows that there are correlated outcomes and spillovers, but may have no idea what the network structure looks like.
Furthermore, this may be in a situation with limited observations of subjects, so there are a few (or finite, from a statistical perspective) clusters available. \cite{viviano2024policy} studies this in both the single and multi-wave settings.

If there are only a few clusters, obviously a global welfare-maximizing treatment assignment cannot be estimated. But even with just a few clusters, if treatments are assigned with enough variation in assignment probabilities, one can estimate whether there is scope for welfare improvement. Specifically, one can estimate the marginal policy effect which is the change in outcome (inclusive of direct and indirect effects) due to a change in the assignment probability and from it see both a direction for welfare improvement but also whether it is worthwhile to conduct subsequent (adaptive) experimentation.

At its core the approach requires that (i) clusters have some weak dependence (e.g., clusters include multiple villages and there are denser links within rather than across villages); (ii) in expectation treatment effects are the same across clusters. The former places restrictions on the underlying network. For instance, the network cannot be too dense (though the exact rate of $o(n^{1/4})$ may be possibly improved) and also the networks are drawn from the same distribution across all clusters.

\subsection{Dealing with Measurement Error}

Measurement error can be a non-trivial problem in the study of networks. There are  several distinct issues.

The first case is where we observe all $n$ nodes but our observed graph $\hat G$ is mis-measured: $\hat G_{ij}\neq G_{ij}$. From the perspective of exposure maps, the worry is that
\[
Y_i(D'_i,D'_{1:n},\hat G) - Y_i(D_i,D_{1:n},\hat G) \neq  Y_i(D'_i,D'_{1:n},G) - Y_i(D_i,D_{1:n}, G).
\]
In essence, since the network is mis-measured, the exposures are mis-coded thereby affecting our estimands. Whether the bias is large depends on the structure of $G$ and the measurement error itself.

There can be many reasons as to why $\hat G \neq G$: survey fatigue / recall bias is one such example whereby the respondent tires of a lengthy survey and therefore does not list all of their links. Other times there may be obvious patterns: the researcher samples $m \leq n$ nodes and for each node $i$ elicits $G_{i,\cdot}$. This gives a mis-measured network $\hat G$ with entries $\hat G_{jk}=0$ if both $j,k$ are not sampled. However, in this case, the mechanics of the missingness can be modeled. A distinct example, where mismeasurement is obvious but the knowledge of the mismeasurement is hard to correct is top-coding: when a respondent is asked to list up to $k$ friends, if they typically have more than $k$ friends then they are coded as having degree $k$ nonetheless.
It could also be that the researcher only has access to some media that presents an incomplete picture of the network.  These are just a few of the possible sources of measurement error.

Whether the fact that $\hat G \neq G$ matters depends on the impact on the exposure maps. Consider an example wherein a study is about a hypothesis that only involves a subject's out-degree. It is quite easy because with a random sample one can actually, discover a perhaps biased version of the degree distribution itself, but the measurement error issues are fairly standard. The sampling is well-understood and so correcting the bias in expectation just requires calculating the appropriate reweightings.
The more challenging sorts of measurement error arise when one needs more detailed information about the interaction patterns, and links and nodes are not all present.
Here there is no easy off-the-shelf remedy, but instead one has to model the noise. At that point one can either proceed by trying to study the robustness of one's estimates (via either frequentist or Bayesian methods) or trying to recover a consistent estimator nonetheless (there are many works on this topic, see, e.g., \cite{chandrasekharl2010,chang2022estimation,lewbel2024ignoring}).

\subsection{Panels}
Thus far we've only briefly discussed panel data, and it is clear panel data can provide considerable value. From an econometric perspective panel data allow a researcher to deal with unobserved, fixed latent parameters.
Here we discuss some examples where experimentalists collected dynamic network data and the different ways in that information was helpful.

\cite{mobiustreasure} are interested in studying a social learning model, which naturally involves multiple waves of conversations and belief updating.  This is naturally a dynamic process and thus naturally suited for collecting information on dynamic conversation networks. They examine how people aggregate information that is initially dispersed in social networks and how network structure affects this process, focusing on distinguishing between "tagging" and DeGroot learning models. "Tagging" refers to a Bayesian learning process where individuals pass along labeled signals identifying the origin of each piece of information.  Thus when people hear something twice that is tagged, they realize that it is the same information and can ignore the second hearing of it.   This contrasts with DeGroot learning, where individuals simply repeatedly average the beliefs of their neighbors, potentially leading to multiple-counting of the same information as it circulates through the network. The experiment employs a "treasure hunt" game where 563 Harvard undergraduates search for a prize on campus over three days, using a custom social networking platform to communicate. This design allows researchers to observe how labeled information spreads and how participants collaborate, while testing whether they engage in tagging (passing labeled signals) or DeGroot-style belief averaging.

The study's first key result shows that information aggregation in the network was highly efficient, with a large fraction of participants finding the correct location, demonstrating the power of social learning. The second result highlights the value of dynamic conversation links in predicting individual performance: centrality in the dynamic conversation network significantly predicted better performance in finding the correct answer, while centrality in static networks (such as Facebook friendships) had little explanatory power. This underscores the importance of capturing actual information flow through dynamic interactions, rather than relying solely on pre-existing social structures, for understanding how information spreads and influences outcomes in task-specific scenarios. The authors leverage the rich dynamic data to estimate a structural model of learning, allowing them to distinguish between tagging and DeGroot learning processes and quantify the efficiency of information aggregation in the network.

As mentioned above, \cite{banerjee2024changes} examines how the introduction of formal credit markets, particularly microfinance, affects the structure of informal social networks in developing economies. The authors investigate both direct effects on credit recipients and indirect effects on overall network structure.
The study uses data from two distinct settings in India, employing different data collection methods. In Karnataka, they collect network data from 75 rural villages, mapping out the social networks explicitly. This rich dataset allows for detailed analysis of network structures and changes.  In contrast, in the second setting of Hyderabad, they use Aggregated Relational Data (ARD) in 102 urban neighborhoods. ARD involves asking individuals about the characteristics of their connections rather than mapping the full network, providing a less detailed but more scalable approach to network measurement.

High (H) types are defined as individuals who based on their demographics and a machine learning model are estimated to be more likely to take out a microcredit loan. Low (L) types are those classified as less likely to take out loans. This classification is crucial for understanding differential impacts of microfinance on network structure.
One main finding, consistent across both settings despite different data collection methods, is that microfinance leads to a reduction in network density. Surprisingly, the largest reductions occur in links between low types and triangles of all low types, rather than in connections involving high types who directly access credit. This is counterintuitive because one might expect that it would be H types, who gain access to formal credit, who have less need for information relationships.  Indeed, relationships involving H types disappear, but relationships between (groups) of L types that do not involve any H types disappear as well and at a level that is as large.
This highlights non-local, general equilibrium effects on network structure. The introduction of microfinance doesn't just affect direct recipients but reshapes the entire network. The consistency of findings between the full network data in Karnataka and the ARD in Hyderabad strengthens the robustness of these conclusions, suggesting that the observed effects are not artifacts of a particular data collection method. It demonstrates that formal credit markets can have far-reaching impacts on social structure, altering informal risk-sharing arrangements even among those not directly accessing the new financial services, and that these effects are detectable using different network measurement approaches.

Both studies underscore the value of panel network data in understanding social and economic processes. While static networks provide context, panel data reveal how networks evolve in response to interventions or environmental changes. \cite{banerjee2024changes} showed microfinance-induced network restructuring, while \cite{mobiustreasure} found panel links more predictive of information aggregation performance than static measures. These studies demonstrate that panel network data can be very helpful in experiments for comprehensively understanding network function and evolution.


\section{Lab Experiments}\label{sec:lab-experiments}

Let us now return to the discussion of lab experiments and discuss their design in more detail.

There are three essential design elements for conducting lab experiments with networks.

The first is the network structure.  This is now no longer something that the researcher is simply trying to measure, but now it becomes a design variable.   Different network topologies give rise to different distributions over outcomes, so researchers need to be strategic in choosing network shapes for lab experiments.  This involves carefully choosing network structures so that comparing across them allows for identification of the key effects.  The size of the networks and how subjects are matched also impacts the power for testing any given question.

The second design element concerns what the subjects know about the network and the incentives of other subjects.  This also depends on the type of hypothesis being tested.
Knowledge of the social structure includes many things, but at least two are often important. First, in settings with nontrivial spillovers and where those extend beyond a subject's immediate neighborhood, their beliefs about the distribution over networks can matter, depending on what they're able to observe, especially in a dynamic experiment.   This lead to complications, since giving subjects only partial information means that then the researcher might have to infer what kinds of beliefs subjects might have, what they can learn during the experiment, and how that influences behavior (see also Chapter 3).
Second, there are questions of what subjects know about other participants' identities and incentives, etc. 
By varying non-anonymity, for instance in lab-in-the-field settings the researcher is projecting the outside network relationships into the lab. This could very well be a feature, not a bug, and may be the object of measurement itself.

The third design element is the scope of behavior, especially with ``lab in the field settings''. The researcher can extend an experiment outside the lab, where A has to send a message to B by a certain date. The third aspect relates to the nature of the costs associated with activating links in the real world. Lab-in-the-field experiments offer a powerful approach to expanding this scope of behavior. By conducting experiments in participants' natural environments, researchers can incorporate real-world social ties, cultural contexts, and economic stakes. This approach allows for the activation of genuine network links and the observation of behaviors that more closely mirror real-life decision-making processes. For instance, experiments conducted in villages or local communities demonstrate how lab-in-the-field methods can capture the nuances of social capital and network-mediated behaviors in ways that pure lab experiments cannot. This methodology enables researchers to bridge the gap between controlled experimental settings and the complexities of real-world social and economic interactions, providing insights that are both theoretically rigorous and practically relevant.

\subsection{Designing Networks in Lab Experiments}

Networked lab experiments present their own challenges.  In particular, any given session of subjects who are all on the same network (usually limited in size by recruiting and/or lab constraints, even if done online), generally have nontrivially correlated outcomes.
Thus the ability to have quasi-independent observations within the network (which is possible in the field) is often precluded in lab settings.  Thus, the unit of analysis often has to be a network of participants, and then one needs many of those networks of participants to have a powerful enough lens to distinguish behavior.   Clever rematching or repositioning of subjects, coupled with careful accounting for interdependence, can improve the power of such analyses substantially and is an important considering in lab experiment design.
It can also be that comparison across subjects in the same network but with different positions can allow one to test a given hypothesis, even though their behaviors might be correlated, since their relative behaviors could be used to test a hypothesis.

The design of network structures in experimental studies thus plays a crucial role in advancing our understanding of social and economic behavior. Two key approaches have proven particularly effective: isolating specific network properties and differentiating between competing theoretical models. These approaches allow researchers to create powerful experimental paradigms to test hypotheses about various social and economic phenomena in networked environments.

Let us begin with some examples of varying network structures allows one to test a theory or distinguish between multiple theories.
\cite{choigk2004,choigk2012} examined how social learning worked in several variations on three-person networks, where people make repeated choices.  The variation of the network changed the sort of inferences that people had to make in order to learn.  For instance, in a complete network, everyone sees everyone else directly, and so there is no need to infer some indirect information via how one's neighbors' behaviors change over time, which then encodes what they have learned.  They find that even in the simplest settings there are some failures of learning, but also that the failures become more pronounced as the network changes so that more indirect inferences have to be made and where changes in others' behaviors have to be observed in order to learn.

\citet{chandrasekhar2020testing} carefully selected network structures to distinguish between the DeGroot and Bayesian models of social learning. They chose variations on 7-node networks that generated divergent predictions under these competing learning models.  The authors derived explicit theoretical predictions for each network and used simulations to verify that these structures would indeed produce distinguishable outcomes.   The combination of networks allowed them to reject the hypothesis of Bayesian learning and show that behavior was more consistent with DeGroot learning.

\citet{centola2010spread} was interested in comparing simple and complex diffusion, where simple is a situation in which a person just needs one contact with someone who is informed/infected to be informed/infected themselves, where complex might require several contacts or friends who have become informed/infected.  Complex diffusion requires networks that have more local closure or cohesion (e.g., more triangles that overlap) while simple diffusion can take place even on trees.  He thus compared clustered-lattice networks to random networks in studying health behavior diffusion. The latter was constructed by essentially rewiring the former, therefore preserving the degree structure. This design allowed for a direct test of whether the diffusion appeared to be better described by simple or complex contagion.

\cite{charness2007bargaining} varied network structure to test a theory of how the gains from trade are split when people bargain and there is a network of potential outside options (different people that a person may trade with, but they also may have other people they can trade with, etc.).  By including some nontrivial network structures, outside options were varied in ways that led to different predictions about gains from trade.   The ways in which the network structure influences equilibrium are intricate and nonobvious, yet the experiment provided a robust test of the theory and found substantial support for the theoretical predictions.  Even though people might not be thinking in terms of equilibrium, the results were consistent with the different network structures naturally guiding subjects' incentives.

\cite{charness2014experimental} studied equilibrium selection in a series of experiments on networks. The games varied whether actions are strategic substitutes or complements and also varied whether the agents know the structure of the network in which they play. This allowed them to show that in cases of a unique equilibrium, individuals behave in accordance with theory; and that in cases with multiple equilibria, one can rather consistently predict equilibrium selection via network features.

\citet{kearns2006experimental} used six different network structures to examine collective problem-solving (solving a coloring problem). Their use of preferential attachment graphs, cycles with random chords, and small-world networks allowed them to study how properties like degree distribution and clustering coefficient impact group performance.  Preferential attachment graphs (which have the thickest degree tails) made solving the problem the hardest whereas the small worlds graph made it the easiest.

\cite{kearnsjtw2009} also used varied networks to see how that affected groups to coordinate on an outcome (all choosing the same color), when they can only see the actions of their neighbors.  Variations in the graph again showed how the network affected outcomes, with more asymmetric networks actually helped coordination, as having most people see the same (small) group of individuals helped in coordination, compared to more random or split graphs in which there was less overlap in observation.

These studies, among others,  collectively demonstrate the power of thoughtful network design in experimental research. By carefully selecting network structures to isolate specific properties or distinguish between competing theories, researchers can create robust experimental paradigms that provide sharp conclusions. This approach has been invaluable in advancing our understanding of how network structure shapes behavior across various social and economic contexts.

The deliberate choice of network structures remains a key tool in the methodological toolkit of network economics. It allows for examination of specific aspects of social interaction and collective behavior, contributing to a deeper understanding of the intricate relationship between network topology and human behavior. Future research in this field will undoubtedly continue to leverage these design principles, pushing the boundaries of knowledge about social and economic dynamics in networked environments.

\subsection{What do subjects know?}

\subsubsection{Network Structure}

When designing a network experiment, a key decision for the experimentalist is determining what information about network structure to provide to participants. This choice shapes subjects' decision-making environment and the questions that can be addressed.

The experimentalist's decision should be guided by two primary considerations:

\begin{enumerate}
    \item The specific economic question being investigated.
    \item The real-world setting the experiment aims to emulate.
\end{enumerate}

For instance, if the goal is to test a theoretical model that requires complete network knowledge, providing full information might be appropriate, or at least comparing how subjects behave when they are in the setting of the model and when they know less.
Conversely, if the research aims to understand behavior in settings where individuals have limited network visibility, as they often due in applications, restricting information could be more suitable.

Given these considerations, experimentalists typically choose from the following information structures:

\begin{enumerate}
    \item Complete Information: Subjects are given full knowledge of the network. This design is useful for testing theories that assume perfect information or for isolating the effects of network position on behavior.  \citet{choigk2004,choi2009social} employ this approach to study information aggregation.  In those experiments, having subjects know the structure of the network is important even if they only observe behaviors of their neighbors, since how they should be updating beliefs under the theory depends on what they know about the network. \citet{chandrasekhar2020testing} use a design of complete network information to compare models of social learning, revealing that subjects tend to employ simple heuristics rather than Bayesian updating even when the network is known to the subjects.
 \citet{charness2014experimental} use complete information in some treatments to study how network knowledge affects equilibrium selection in coordination games.

    \item Local Information: Subjects only know their immediate connections and are uncertain about others'. This setup often mirrors real-world limitations in network knowledge and is valuable for studying local learning processes. For example, see \citet{kearns2006experimental,kearnsjtw2009,mueller2015social}.

    \item Varied Information: Some experiments compare treatments with different levels of network information. This approach, as reviewed by \citet{choi2016handbook}, allows for direct testing of how information structures affect economic outcomes.
\end{enumerate}

The choice among designs should be informed not only by the theory being test, but also by field evidence on how individuals perceive and navigate their social networks. For example, \citet{breza2018networks} document significant discrepancies between perceived and actual network structures in the field, suggesting that experiments with incomplete or local information might better approximate real-world conditions.
Ultimately, the experimentalist's decision about network information provision shapes the external validity of the results and the types of economic insights that can be gained. By carefully considering this design element, researchers can create experiments that not only test theoretical predictions but also provide insights into how individuals behave in networked environments under realistic information constraints.

\subsubsection{Non-Anonymity}

The study of social and economic networks has been significantly enriched by the introduction of non-anonymous experimental design, which are often set in lab-in-the-field experiments. These designs, which allow participants to know each other's identities, offer several key advantages over traditional anonymous laboratory experiments.

First and foremost, non-anonymous designs enhance the external validity of our findings. By allowing participants to interact with known peers, we create an environment that more closely mirrors real-world social and economic interactions. This is particularly crucial when studying phenomena such as contract enforcement in the absence of formal institutions, as demonstrated in our work \citep{chandrasekhar2018social}. That study showed that network structure and people's positions in the network significantly influence individuals' propensity to honor informal contracts, a finding that would not have been possible in an anonymous setting.

The known-identity approach also allows us to delve deeper into the mechanics of social distance and its effects on economic behavior. \citet{goeree20101} leveraged this approach to establish their ``1/d law of giving,'' showing how prosocial behavior declines with social distance. This finding has important implications for our understanding of how network structure influences resource allocation and risk-sharing behaviors.

Moreover, non-anonymous designs enable us to study the formation and evolution of trust and reciprocity within networks. \citet{riedl2002exclusion} demonstrated how individuals strategically form and break links based on past interactions, a process that is fundamentally tied to the identifiability of network partners, and has important implications for the evolution of cooperation.

\cite{jacksonx2014} show how people's choice of strategies in an asymmetric coordination game depends on their country of origin, and their ability to forecast their partner's action and to coordinate depended on whether they were matched with people from the same or different countries.  Thus, subjects' backgrounds can systematically influence their behaviors (e.g., see also \cite{henrichetal2001}), and better understanding of how their interaction shapes their beliefs and behaviors is needed.

Another critical advantage of non-anonymous designs is their ability to capture the influence of network position in high-stakes environments. For instance, \citep{chandrasekhar2018social} study games with stakes as high as 1-2 days' wages where randomly matched participants may vary considerably in relative network position; this random matching then generates variation helping to identify how network position may substitute for formal contracts. Both \cite{leider2009directed} and
 \citet{ligon2012motives} parse out motives for giving---altruism, reciprocity, etc.---with a view to network position, which is possible by playing non-anonymous games (at least in certain arms of the study) paired with network data.

\subsection{Scope of behavior}

A key advantage of lab-in-the-field experiments is the ability to leverage existing social networks. \citet{chandrasekhar2018social} exemplify this approach, conducting their study in rural Indian villages utilizing pre-existing social ties to examine informal contract enforcement. Similarly, \citet{ligon2012motives} exploit information on actual social networks in Paraguayan villages to study sharing motives. By activating genuine social connections, these studies capture the nuanced effects of longstanding relationships and accumulated social capital on economic decision-making -- aspects that are challenging to replicate in traditional lab settings.

Moreover, lab-in-the-field experiments often involve higher-stakes, more contextually relevant decisions. While \citet{leider2009directed} worked with college students, they framed their experiments within real social networks, enhancing the salience of decisions. \citet{caria2020expectation} pushed this further, embedding their experiment in the context of job search and professional networking in urban Tanzania. This approach not only bolsters external validity but also potentially elicits more authentic behaviors from participants who perceive the stakes as personally meaningful.

The synthesis of experimental control and real-world elements in lab-in-the-field studies enables researchers to decompose complex social and economic phenomena that might be conflated in pure lab or field settings. \citet{leider2009directed} disentangle baseline altruism, directed altruism, and enforced reciprocity while accounting for real social relationships among college students. \citet{ligon2012motives} perform a similar decomposition of sharing motives in village networks. This approach of isolating mechanisms while maintaining real-world relevance is a hallmark of lab-in-the-field experiments. It facilitates a more granular understanding of how social networks mediate economic behavior, trust, cooperation, and risk-sharing across various contexts. \citet{caria2020expectation} further exemplify this by manipulating expectations about future interactions and observing effects on actual networking behaviors, demonstrating how lab-in-the-field methods can test theoretical predictions in situ. Collectively, these studies underscore the potential of lab-in-the-field experiments to generate insights that are both theoretically rigorous and practically relevant, informing both academic discourse and policy design in network economics.

\subsection{Experiments on Network Formation}

Given the importance of networks in determining behaviors, understanding how networks are formed becomes of great importance (e.g., see the discussion and references in \cite{jackson2003,jackson2005survey,jackson2014}).

As discussed above, there are field settings in which one can test some models of network formation.  However, the ability to observe and control all the relevant factors can be quite challenging, especially when they involve questions of individual rationality and behavior in forming relationships, when that process might be difficult to observe in the wild.

As such, there are many experiments that leverage laboratory control to test theories of how networks form, including
\cite{charnessj2007,corbae2008experiments,comola2014testing,falk2012s,goeree2009search,corten2010co,kearns2012behavioral,rong2015growing,kirchsteiger2016limited,teteryatnikova2020myopic,horvath2024network}.

For example, if one wants to examine the extent to which people are forward looking as to how the future network will evolve when they form their own relationships, that can be especially challenging in the wild since there are many unobserved factors that could influence people's beliefs about the future and their patience, etc.   In the lab, these can all be controlled.  For example, by using a dynamic lab setting \cite{kirchsteiger2016limited} carefully test the extent to which subjects are forward looking vs myopic in forming relationships (e.g., how do they account for how the future network will evolve and how that influences their payoffs), finding subject heterogeneity and behavior often between the two extremes.

\section{Natural Experiments}

Another form of experiment is one in which ``Nature'' has somehow done something that can substitute for the sort of randomization that a researcher would like to do.  The advantage of this approach is that it often operates at a scale that a researcher cannot, or affects variables that a researcher cannot.

Usually in natural experiments the randomization that is being done by nature has to do with the treatment status.   For instance, some population gets exposed to a program while another does not.   For instance, in \cite{banerjee2024changes} we explored what happens to networks and consumption smoothing when some villages get microfinance and other villages do not. 
In \cite{araln2017} the authors use random weather events to vary the behaviors of a subject's friends on an exercise app and then see how the subject reacts to different behaviors of their friends (which are now randomized independently of other things that might influence the subject's decision of whether to exercise).

When it comes to networks, however, another type of natural experiment that can be extremely useful is that different subjects may be placed in different network positions.  Thus, it is the network itself that is being randomly controlled by nature.   This can be incredibly useful, as controlling networks on a large scale and in a natural setting is otherwise usually impossible for a researcher.
A few examples of this sort of approach are \cite{sacerdote2001}, \cite{beaman2012}, and \cite{laschever2011}.   \cite{sacerdote2001} took advantage of random allocation of roommates in a college, \cite{beaman2012} took advantage of random locations of refugees which affected how many people there were from similar backgrounds in their new networks, and \cite{laschever2011} took advantage of the random assignment of drafted subjects to military units to form their friendship networks.  All of these could then see how people's outcomes differ dependent upon whom that are put in contact with---essentially the networks that they form.

Of course, as with any experimental design, one has to worry about the extent to which the thing one would like to (randomly) vary is really being varied completely at random (e.g., see some discussion in \cite{angrist2014perils}).  Is the network really being varied?  Do people have other outside networks upon which they might differentially rely upon and which are not observed?  Is there some sort of selection or imperfect aspect to the randomization?   These are things that are an issue in any experiment, but can become more problematic when it is not the researcher designing and controlling the randomization.  Nonetheless, one can still make some sense of the outcomes, with appropriate cautions and adjustments \citep{angrist1996identification}.


\section{Concluding Thoughts and Synergies across Types of Experiments}\label{sec:discussion}

There exists a clear set of trade-offs between laboratory and field experiments in network economics. Lab experiments offer researchers precise control over network structures, allowing for clean identification of causal mechanisms and high internal validity. They also provide clearer interpretability of results due to their controlled environment. However, this comes at the (usual) costs of potentially limited external validity and real-world relevance. Field experiments, on the other hand, leverage existing social networks and offer higher external validity, capturing the complexity and nuances of real-world interactions. They provide insights directly applicable to policy and practice. Of course,  field experiments often sacrifice some degree of control and interpretability, as there can be nontrivial measurement error and unobserved factors.

In this way the two techniques can serve as useful complements, where the same theories can be tested both in the lab and the field, and in a variety of different contexts.
Thus, we can see how lab, field, and lab-in-the-field experiments all interact and speak to each other. This helps to highlight the trade-offs between controlled lab settings and real-world applications, with lab-in-the-field experiments bridging the gap.

For example, recall \citet{chandrasekhar2020testing}'s social learning experiment.  Such controlled environments enable researchers to isolate specific mechanisms, precisely define network structures, and cleanly test theoretical predictions (here DeGroot versus Bayesian learning, possibly with incomplete information). However, this control comes at the expense of external validity   and potential behavioral artifacts, since it is not immediate that real-world learning would reflect what we see when people make decisions over small decisions in a learning experiment. At the same time, if taken seriously, the conclusions of this lab experiment help us deepen our understanding of diffusion processes.

At the other end of the spectrum,  \citet{banerjee2023selecting} demonstrate the power of large-scale field experiments, implementing an immunization intervention across 40\% of Haryana, India. This approach offers real-world relevance, leverages existing social networks, and provides policy-applicable insights from network theory (by identifying information ambassadors who are likely to be central in the unobserved, real-world network). The trade-off here lies in reduced control over network structures, increased measurement noise due to real-world complexities, and potential lack of interpretability.
Also, even such field experiments have some limitations in external validity, as what works in Haryana might not work in other regions or countries.

Lab-in-the-field experiments serve as a crucial intermediate step, balancing control with context. These studies can utilize actual social ties while imposing controlled information flows, or frame decisions in locally relevant terms while maintaining experimental rigor. This approach facilitates external validity tests of lab findings and explores how real social structures influence behaviors observed in controlled settings.

Network economics, given the complex nature of the subject matter, demands that we attempt to tackle these questions from all angles. Here we see that one can take a research arc and   climb from surgical findings to complex and more messy, real-world applications. It allows for theory refinement, identification of persistent mechanisms across contexts, and development of evidence-based policies. The progression from \citeauthor{chandrasekhar2020testing}'s (\citeyear{chandrasekhar2020testing}) controlled network study to \citeauthor{banerjee2023selecting}'s (\citeyear{banerjee2023selecting}) large-scale immunization experiment provides an example as to how  insights from controlled settings can inform and improve real-world interventions.

\


\bibliographystyle{ecta}
\bibliography{networks}

\end{document}